\documentclass[preprint2]{aastex631}

\usepackage{nccmath}
\usepackage{xcolor}
\usepackage{hyperref}
\usepackage[anythingbreaks]{breakurl}
\usepackage{graphicx}
\newcommand{\degree}{$^\circ$}

\usepackage{soul}
\usepackage{CJK}
\usepackage{tabularx}
\usepackage[shortlabels]{enumitem}
                    \setlist[enumerate, 1]{1\textsuperscript{o}}

\begin{document}
\title{Improving the Automated Coronal Jet Identification with U-NET}

\correspondingauthor{Jiajia Liu}
\email{jiajialiu@ustc.edu.cn}

\author[0000-0003-2569-1840]{Jiajia Liu}
\affiliation{Deep Space Exploration Lab/School of Earth and Space Sciences, University of Science and Technology of China, Hefei, 230026, China}
\affiliation{CAS Key Laboratory of Geospace Environment/CAS Center for Excellence in Comparative Planetology/Mengcheng \\ National Geophysical Observatory, University of Science and Technology of China, Hefei, 230026, China}
\author[0009-0006-4430-7882]{Chunyu Ji}
\affiliation{CAS Key Laboratory of Geospace Environment/CAS Center for Excellence in Comparative Planetology/Mengcheng \\ National Geophysical Observatory, University of Science and Technology of China, Hefei, 230026, China}

\author[0000-0002-8835-3825]{Yimin Wang}
\affiliation{School of Data Science, Qingdao University of Science and Technology, Qingdao, 266100, China}

\author[0000-0002-3606-161X]{Szabolcs So{\'o}s}
\affiliation{Department of Astronomy, E\"{o}tv\"{o}s Lor\'{a}nd University, Budapest, P\'{a}zm\'{a}ny P. s\'{e}t\'{a}ny 1/A, H-1117, Hungary}
\affiliation{Gyula Bay Zolt\'an Solar Observatory (GSO), Hungarian Solar Physics Foundation (HSPF), Pet\H{o}fi t\'er 3., Gyula, H-5700, Hungary}

\author[0000-0002-6683-0205]{Ye Jiang}
\affiliation{School of Information Science and Technology, Qingdao University of Science and Technology, Qingdao, 266100, China}

\author[0000-0003-3439-4127]{Robertus Erd{\'e}lyi}
\affiliation{Solar Physics and Space Plasma Research Centre (SP2RC), School of Mathematics and Statistics, The University of Sheffield, Sheffield, S3 7RH, UK}
\affiliation{Department of Astronomy, E\"{o}tv\"{o}s Lor\'{a}nd University, Budapest, P\'{a}zm\'{a}ny P. s\'{e}t\'{a}ny 1/A, H-1117, Hungary}
\affiliation{Gyula Bay Zolt\'an Solar Observatory (GSO), Hungarian Solar Physics Foundation (HSPF), Pet\H{o}fi t\'er 3., Gyula, H-5700, Hungary}

\author[0000-0002-0049-4798]{M. B. Kors\'os}
\affiliation{University of Sheffield, Department of Automatic Control and Systems Engineering, Amy Johnson Building, Portabello Street, Sheffield, S1 3JD, UK}
\affiliation{Department of Astronomy, E\"{o}tv\"{o}s Lor\'{a}nd University, Budapest, P\'{a}zm\'{a}ny P. s\'{e}t\'{a}ny 1/A, H-1117, Hungary}
\affiliation{Gyula Bay Zolt\'an Solar Observatory (GSO), Hungarian Solar Physics Foundation (HSPF), Pet\H{o}fi t\'er 3., Gyula, H-5700, Hungary}

\author[0000-0002-8887-3919]{Yuming Wang}
\affiliation{Deep Space Exploration Lab/School of Earth and Space Sciences, University of Science and Technology of China, Hefei, 230026, China}
\affiliation{CAS Key Laboratory of Geospace Environment/CAS Center for Excellence in Comparative Planetology/Mengcheng \\ National Geophysical Observatory, University of Science and Technology of China, Hefei, 230026, China}

\begin{abstract}
Coronal jets are one of the most common eruptive activities in the solar atmosphere. They are related to rich physics processes, including but not limited to magnetic reconnection, flaring, instabilities, and plasma heating. Automated identification of off-limb coronal jets has been difficult due to their abundant nature, complex appearance, and relatively small size compared to other features in the corona. In this paper, we present an automated coronal jet identification algorithm (AJIA) that utilizes true and fake jets previously detected by a laborious semi-automated jet detection algorithm (SAJIA, Liu et al. 2023) as the input of an image segmentation neural network U-NET. It is found that AJIA could achieve a much higher (0.81) detecting precision than SAJIA (0.34), meanwhile giving the possibility of whether each pixel in an input image belongs to a jet. We demonstrate that with the aid of artificial neural networks, AJIA could enable fast, accurate, and real-time coronal jet identification from SDO/AIA 304 \AA\ observations, which are essential in studying the collective and long-term behavior of coronal jets and their relation with the solar activity cycles.

\end{abstract}

\section{Introduction} \label{sec:intro}

Jets are abundant in the solar atmosphere. A large amount of jets with various scales and temperatures originating from different locations on-disk or off-limb have been observed using modern telescopes since the first observations of H$\alpha$ surges \citep[``cold jets'',][]{Newton1934} almost one century ago. Based on their different sizes, solar jets are often divided into two categories: small-scale jets and large-scale jets. 

Small-scale jets are usually referred to as spicules. Spicules are further sub-divided as the traditional Secchi-type (also called as type-I) and the ones generated by magnetic reconnections (type-II, which is also referred to as rapid blue/red excursions, RBEs/RREs), while both are usually observed in the chromosphere and transition region \citep[e.g.,][]{Beckers1968, Sterling2000, DePontieu2007, Sekse2012}. 
The importance of small-scale jets is well-known as they are suggested to have substantial contributions to coronal heating and solar wind acceleration \citep[e.g.,][]{He2009, Moore2011, Goodman2012, Samanta2019}. The triggering mechanisms of spicules are complicated, which could involve (combined) effects of small-scale magnetic reconnections \citep[e.g.,][]{DePontieu2007, Samanta2019}, waves \citep[e.g.,][]{Heggland2007, Jess2009, Jess2012, Dey2022}, and vortices/Alfv{\'e}n pulses \citep[e.g.,][]{Liu2019APJ, Liu2019NC, Oxley2020, Battaglia2021, Scalisi2021, Scalisi2021b}.

Large-scale jets have been given different names based on the passbands they are observed in, including white-light jets \citep[e.g.,][]{Filippov2011, KUDRIAVTSEVA2019}, H$\alpha$ surges \citep[e.g.,][]{Brooks2007, Zhelyazkov2015}, UV/EUV jets \citep[e.g.,][]{Liu2015Twin, Chen2017, Liu2019How, ZhangQ2021, Schmieder2022}, and X-ray jets \citep[e.g.,][]{Shibata1992, Cirtain2007}. Although various models have been proposed \citep[e.g.,][]{Shibata1992, Canfield1996, Moore2010, Sterling2015, Pariat2015}, almost all have magnetic reconnections, especially the interchange reconnection between open and closed magnetic field lines, involved as the triggering mechanism of large-scale jets. Besides, they have been widely found to be related to many phenomena at different scales, including rotational motions \citep[e.g.][]{Liu2014How, Raouafi2016, Shen2021}, waves/instabilities \citep[e.g.,][]{Giannios2006, Cirtain2007, Kuridze2016, Bogdanova2017, Zhao2018, Li2023}, blobs \citep[e.g.,][]{Zhang2014, Ni2017, Chen2022}, radio bursts \citep[e.g.,][]{Mulay2016, Hou2023}, ``switchbacks'' in the solar wind \citep[e.g.,][]{Sterling2020, Raouafi2023}, and coronal mass ejections \citep[CMEs, e.g.,][]{Shen2012, Liu2015JetCME, Zheng2016, Chenhe2021}. These have made solar jets one of the most important phenomena that connect small and large scales, lower and higher layers, and flows and waves in the highly magnetized and stratified solar atmosphere.

Owing to their complex observational features and abundant nature, it has been rare to study the statistical and long-term behavior of solar jets using a dataset with a large number of events, although we have now entered an era with a tremendous amount of high-spatial and high-temporal resolution observations of the Sun. \cite{Musset2023} started a citizen science initiative called ``Solar Jet Hunter'' to utilize human resources worldwide in manually identifying coronal jets observed in the Atmospheric Imaging Assembly \citep[AIA, ][]{Lemen2012} 304 \AA\ passband onboard the Solar Dynamics Observatory \citep[SDO,][]{Pesnell2012}. Although more than 800 coronal jets have been reported, this approach suffers from some shortcomings, including the low efficiency of manual identification and the inconsistency of the criteria between different individuals, where the latter could pose unknown bias when statistical analysis is performed. 

\begin{figure*}[t!]
\centering
\includegraphics[width=0.9\hsize]{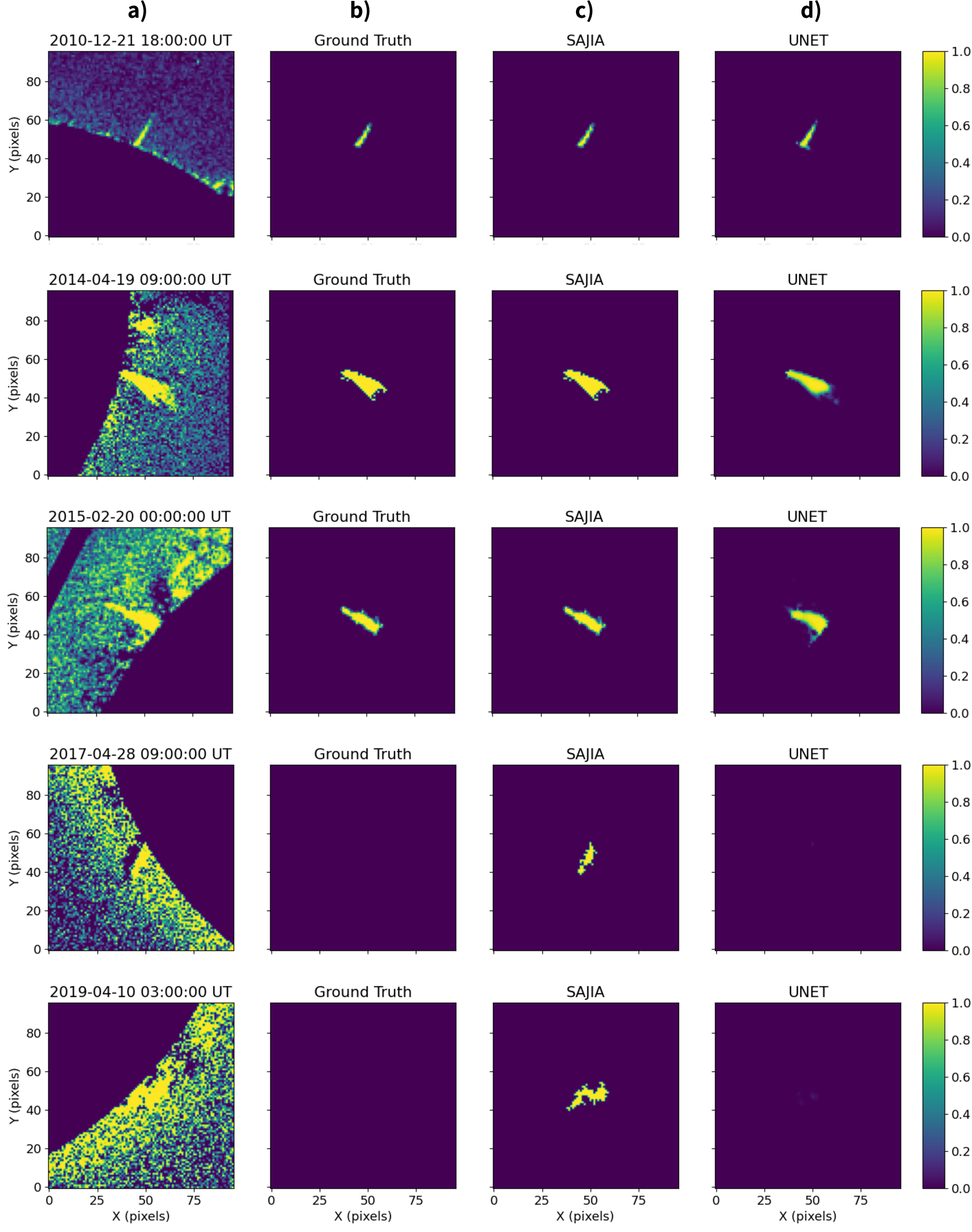}
\caption{\textbf{Examples of coronal jets and detection results.} Images in panel a) are patches of SDO/AIA 304 \AA\ observations with a size of 96$\times$96 pix$^2$. Panel b) lists the corresponding ground truths where yellow colors denote pixels belonging to jets. Panel c) shows detection results by the semi-automated jet detection algorithm developed by \cite{Liu2023Power}. Panel d) are the jet detection results by the automated jet identification algorithm (AJIA) proposed in this paper.}
\label{fig_examples}
\end{figure*}

To facilitate more systematic studies of off-limb coronal jets with less human biases, \cite{Liu2023Power} developed a semi-automated jet identification algorithm (SAJIA) based on applying traditional computer vision techniques to SDO/AIA 304 \AA\ observations. More than 1200 coronal jets were detected by applying SAJIA to SDO/AIA 304 \AA\ observations obtained from 2010 to 2020. A power-law distribution of the jets' thermal energy was found to be highly consistent with those of micro-flares, indicating that they should result from the same nonlinear statistics of scale-free processes. This result was also supported by the first coronal jet butterfly diagram, which is usually seen in the migration of sunspots during solar activity cycles. By doubling the number of observations and extending them to the end of 2021, \cite{Soos2024} expanded the dataset to more than 2700 coronal jets and found some intriguing oscillatory behaviors from their spatial-temporal distributions. It is worth noting that many of these detected jets are in polar regions. Previous studies suggest that solar jets at various scales \citep[e.g.,][]{Chandrashekhar2014,Chitta2023,Uritsky2023} could contribute to energizing the solar wind. The above dataset would enable further such studies from a statistical perspective.

However, it should be noted that the automated identification part in SAJIA has a relatively low precision ($\sim$0.34) and suffers from the CCD degradation of the AIA instrument \citep{Santos2021, Liu2023Power, Soos2024}. A way to address the above issue was to check the identification results manually to eliminate fake jets. The above process was time-consuming and prevented SAJIA from being deployed for real-time jet detection. This paper presents the automated jet identification algorithm (AJIA) with the U-NET neural network \citep{Ronneberger2015}. We demonstrate that the average precision of AJIA is above 0.8, which enables a more accurate coronal jet detection. The paper is organized as follows: the dataset is described in Sect.~\ref{sec:data} with the model and training process detailed in Sect.~\ref{sec:methods}. Results are presented in Sect.~\ref{sec:result}, before the conclusions and discussions in Sect.~\ref{sec:conc}


\section{Data} \label{sec:data}
True and fake jets detected by SAJIA are used as the input of the U-NET model to be detailed in the next section. The method of SAJIA \citep{Liu2023Power} is briefly recapped as follows:

\begin{itemize}
    \item For a  given SDO/AIA 304 \AA\ image, a background is constructed using four images obtained on the same day and then subtracted from the given image.
    \item The solar disk (with a radius of 1.02 solar radii) of the background-removed image is masked as we only detect off-limb jets.
    \item The masked image (4096$\times$4096 pix$^2$) is downgraded to 512$\times$512 pix$^2$ to reduce the computational power needed.
    \item The downgraded image is then normalized and binarised with given thresholds.
    \item The Douglas-Peucker algorithm \citep{Douglas1973} is employed to determine the shape of bright features in the image and yield candidate polygons.
    \item Polygons with four edges, inclination angles less than 60\degree\, and aspect ratios greater than 1.5 are kept as jet candidates.
    \item Each candidate is manually checked to determine whether it is a true or fake jet.
\end{itemize}

By applying the above processes to SDO/AIA 304 \AA\ observations from 2010-06-01 to 2021-12-31 with six images per day at a cadence of 3 hours from 00 UT, 7890 jet candidates were detected \citep{Liu2023Power, Soos2024}. Among all the candidates, 2704 are found to be true jets, and 5186 are fake ones, resulting in a precision of $2704/7890 \approx 0.34$. Initially, full-disk images that contain the above jet candidates were used to build the input of the U-NET model. However, it resulted in poor performance, with the model generating all off-limb bright features but not focusing on jets. This result was unsurprising as jets are relatively small in the full-disk observations, and other bright features would have introduced many distractions to the neural network model.

\begin{figure*}[htb]
\centering
\includegraphics[width=\hsize]{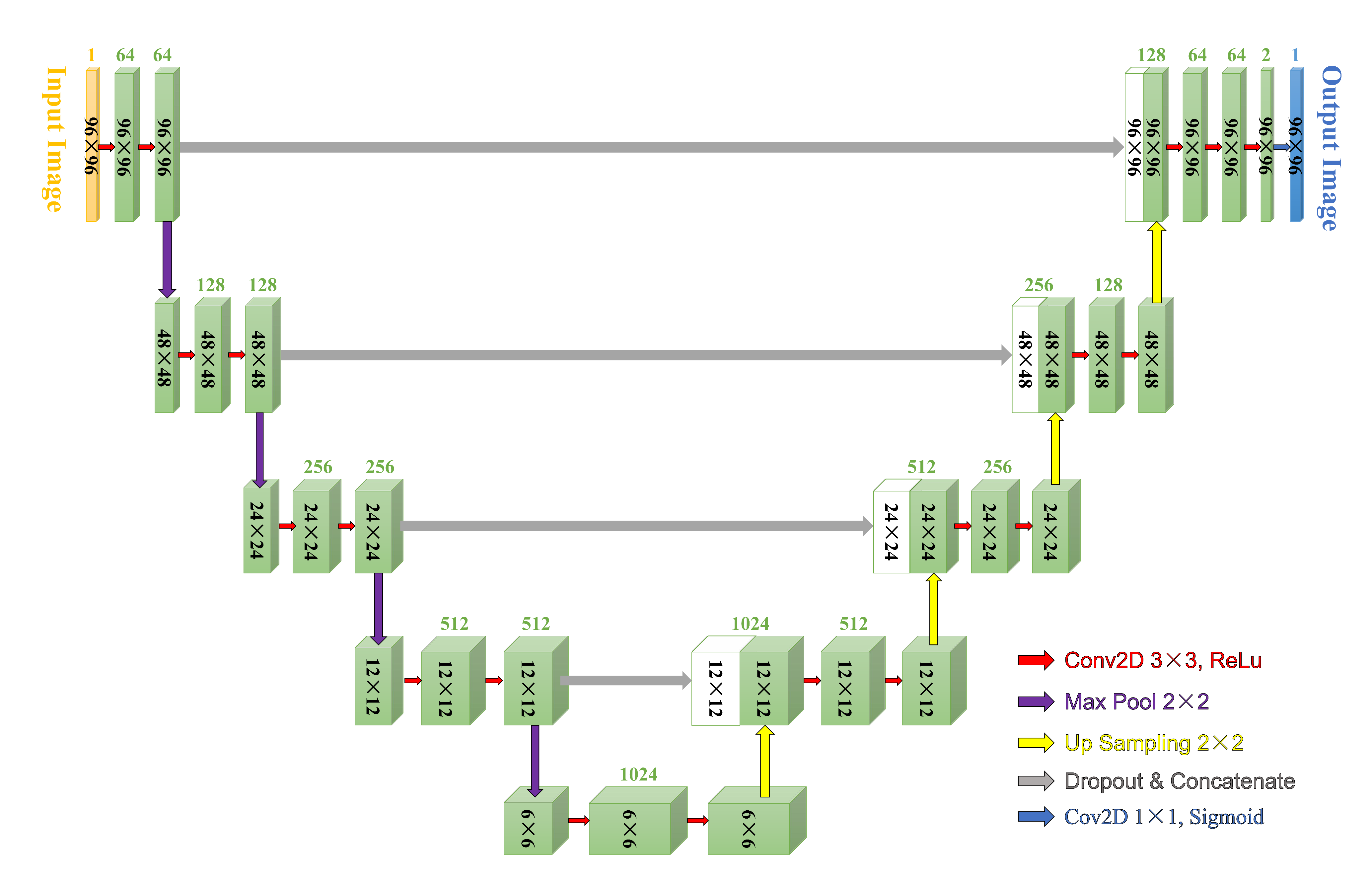}
\caption{\textbf{Architecture of U-NET.} This cartoon is adopted from \cite{Ronneberger2015}. See Sect.~\ref{sec:methods} for a detailed description of the U-NET architecture.}
\label{fig_unet}
\end{figure*}

Considering that the largest detected jet has a length of approximately 70 pixels, small patches of 96$\times$96 pix$^2$ centered at each jet candidate are then extracted from the masked observations (see the first three steps in SAJIA described above). This particular size of patches could minimize the appearance of non-jet features in the images while ensuring that one single jet would not be cut into several patches. These patches are then normalized to [0, 1] with a threshold of 5, which was determined via trial and error. Some examples of these patches are shown in Figure~\ref{fig_examples} a), where the first three rows are true jets, and the last two contain fake jets. Patches with the same size are generated to serve as the ground truth (labels) of the neural network, with pixels covered by true jets set to 1 and all other pixels set to 0. Figure~\ref{fig_examples} b) are the ground truth of the corresponding observations in panel a). SAJIA detections are depicted in panel c). 

The resulting dataset contains 2704 (5186) pairs of image and label patches of true (fake) jets. These true jets have projected lengths in the plane of the sky from more than 10 Mm to about 330 Mm. All jets are divided into two parts: 80\% into the train set and 20\% into the validation set. However, the dataset is imbalanced as there are 91.7\% more fake jets than true jets. To solve this problem, new images and labels are generated by randomly flipping and rotating (between $\pm 0.4\pi$, big enough while less than $0.5\pi$ above which many parts of the images would be cropped) the original images and labels of true jets. The above data augmentation is performed separately in the train and test sets to avoid possible data leakage. After the data augmentation, the dataset is balanced with 5186 fake jets and 5408 true jets. The final dataset has 8474 (2120) pairs of images and labels in the train (test) set. 

\section{Model and Training} \label{sec:methods}

\begin{figure}[htb]
\begin{center}
\includegraphics[width=\hsize]{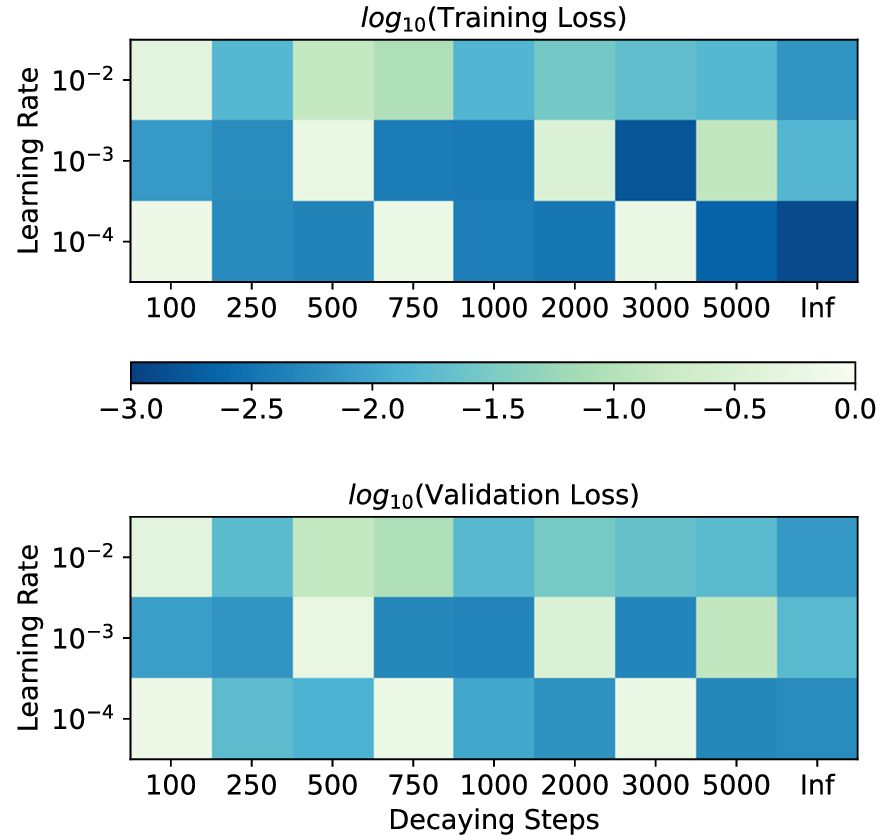}
\caption{\textbf{Model losses under different learning rates}. The upper (lower) panel shows the logarithm of the training (validation) losses for different combinations of learning rates and decaying steps. A decaying step of $Inf$ means the learning rate is fixed without any decaying.} 
\label{fig_loss}
\end{center}
\end{figure}

The U-NET convolutional neural network architecture (see Fig.~\ref{fig_unet}) was initially proposed by \cite{Ronneberger2015} for biomedical image segmentation purposes. It was applied to transmitted light microscopy images and won the ISBI cell tracking challenge 2015 \citep{Ronneberger2015}. U-NET was then modified and successfully applied for different kinds of image segmentation purposes, including 3D image segmentation and road segmentation \citep[e.g.,][]{Minaee2021}.

U-NET contains several convolutional layers with different filter sizes. In the first step, the input image with a size of $96\times96\times1$ (yellow block in Fig~\ref{fig_unet}) is taken into two convolutional layers, each having 64 filters with a kernel size of $3\times3$. In each layer, the Rectified Linear Unit (ReLU) activation function is used after each convolutional operation, where $ReLU(x)=max(0, x)$. The resulting image after the first step has a size of $96\times96\times64$. This image is then down-sampled to $48\times48\times64$ by a max pooling operation (purple arrows in Fig.~\ref{fig_unet}), where only the maximum value in every $2\times2$ region in the image is kept, and all other pixels are discarded. The image is then taken into the next step, which contains two convolutional layers but doubles the number of filters (128).

The above process is repeated until the image size is down-sampled to $6\times6$ but with 1024 filters. Then, a reverse series of operations of the above process is performed to up-sample (yellow arrows in Fig.~\ref{fig_unet}) the image until it again has a size of $96\times96\times64$. Two extra convolutional layers with 2 and 1 filters are used to generate the final output image (blue block in Fig.~\ref{fig_unet}). This unique convolutional neural network is named ``U-NET'' as its architecture resembles the letter U (Fig.~\ref{fig_unet}). The left (right) part of U-NET is usually called the encoder (decoder). Layers in the encoder are skip-connected with layers in the decoder (grey arrows in Fig.~\ref{fig_unet}). These skip connections remind U-NET of the fine details learned in the encoder that could be used to construct images in the decoder. It has been found particularly effective and successful in image segmentation as its contracting path (down-sampling) can capture the context of an image, and its symmetric expanding path (up-sampling) can enable precise localization \citep{Ronneberger2015}. The loss function of U-NET is set to be the binary cross-entropy loss, which is defined as follows:

\begin{equation}
\label{eq1}
\centering
log(L) = \frac{1}{N}\sum_{i=1}^N \left[ y_i log(\hat{y_i}) + (1-y_i) log(1-\hat{y_i})\right],
\end{equation}

\noindent where, $L$ is the loss. $N$ is the number of pixels in each image. $y_i$ is the value (0 or 1) of each pixel in the label, and $\hat{y_i}$ is the corresponding prediction value (0 to 1).

\begin{figure}[t!]
\centering
\includegraphics[width=\hsize]{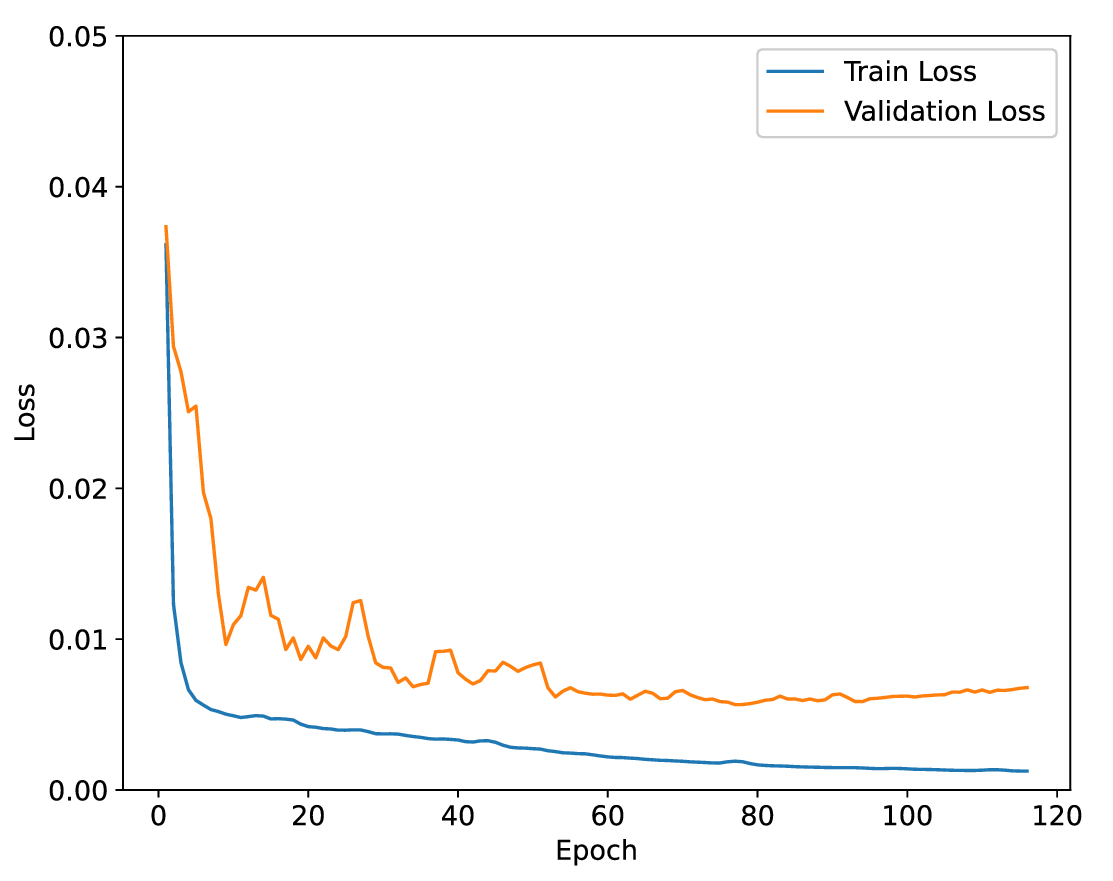}
\caption{\textbf{Training history with a fixed learning rate of $10^{-4}$.} The blue (orange) curve is the evolution of the training (validation) loss.}
\label{fig_history}
\end{figure}

All 8474 pairs of images and labels in the train set were taken into the above U-NET neural network to train a jet identification model. Considering the capacity of the GPU (Nvidia GeForce GTX 4090 with a RAM of 24 GB), the batch size was set to be 256. Another vital hyper-parameter during training is the learning rate. The learning rate determines how much the model weights are updated in response to the loss of each batch. A too-big learning rate will result in the model skipping the minimum of the loss function and making it hard to converge, while a too-small learning rate will probably trap the model in a local minimum of the loss function. A common practice is to set a relatively large learning rate at the beginning of the training and decrease it along a given function at certain steps. In the case of our model, a cosine decay function is used to avoid the learning rate decreasing too fast \citep[see, e.g., ][ for more details]{Loshchilov2016}.

Figure~\ref{fig_loss} depicts the losses obtained in the train (upper panel) and validation sets (lower panel) with different initial learning rates and decaying steps. An ``infinite'' decaying step represents a fixed learning rate without decay during training. It can be seen from Figure~\ref{fig_loss} that the minimum losses (in the order of $\sim10^{-3}$) in both of the train and validation sets can be achieved when the learning rate is fixed at $10^{-4}$.

\section{Results} \label{sec:result}

\subsection{Model selection}
The blue curve in Figure~\ref{fig_history} is the evolution of the training loss, with a fixed learning rate of $10^{-4}$. It is seen that the training loss decreases as the number of training epochs grows and reaches its minimum of $1.2\times10^{-3}$ at 117 epochs. At 117 epochs, the validation loss (orange curve in Fig.~\ref{fig_history} a) is $5.5\times10^{-3}$, which is also around its minimum. Two other commonly used parameters to measure the performance of image segmentation tasks are the mean average precision (mAP) and the mean intersection-over-union (mIoU). Here, we report that, at 117 epochs, the trained model has an mAP of 0.87 and an mIoU of 0.50 for the training set, an mAP of 0.58, and an mIoU of 0.50 for the validation set.

\begin{figure}[htb]
\begin{center}
\includegraphics[width=\hsize]{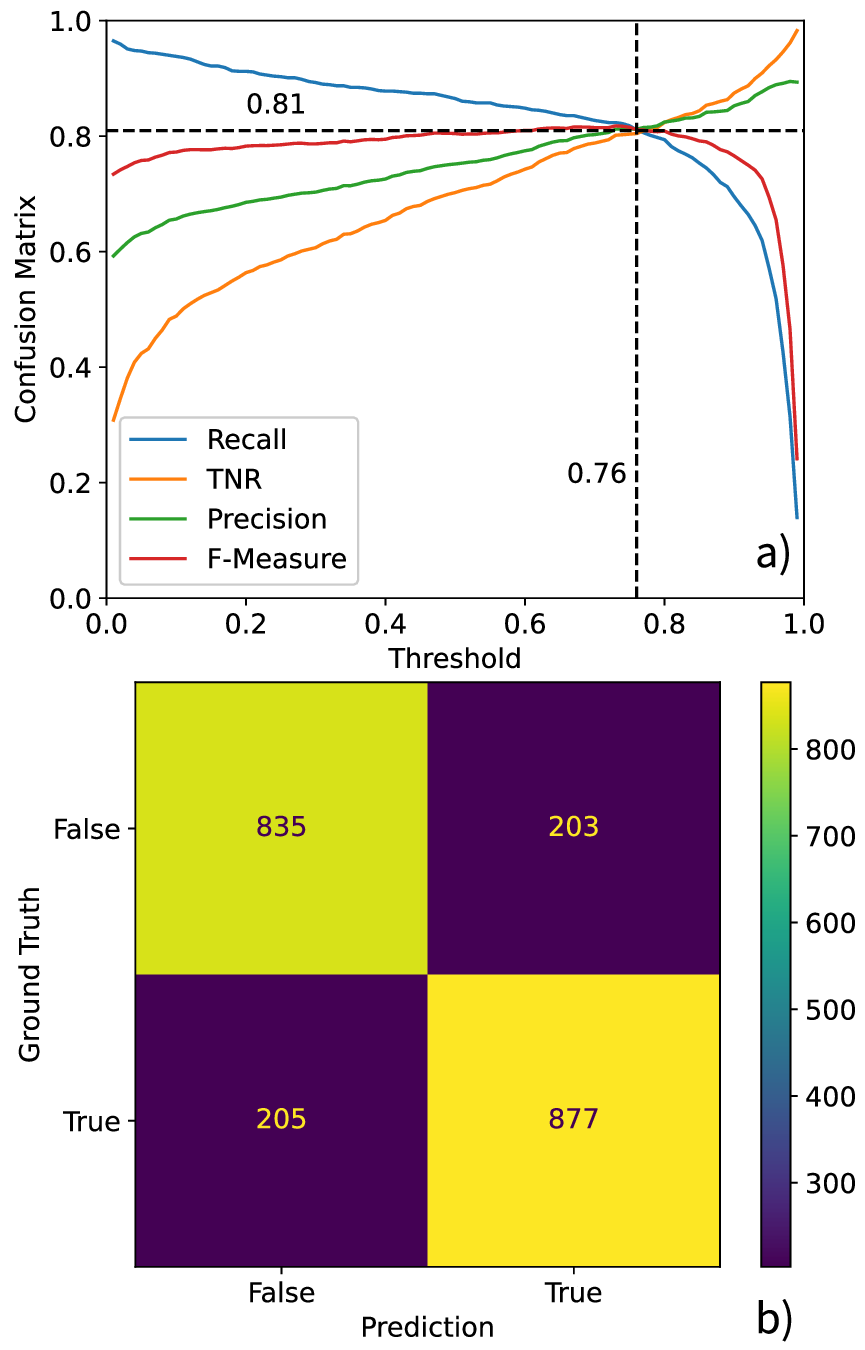}
\caption{\textbf{Confusion matrix of the trained U-NET model.} Panel a) shows how the recall (blue curve), TNR (orange curve), precision (green curve), and F-measure (red curve) changes with different thresholds. See Equation~\ref{eq1} for the definitions of the above measurements. Panel b) is the distribution of true and fake jets in the input labels and predictions made by AJIA.} 
\label{fig_matrix}
\end{center}
\end{figure}

Based on the above observations, we use the model trained at 117 epochs as the final coronal jet identification model. This model could take the $96\times96$ patches of the SDO/AIA 304 \AA\ observations as its input and automatically identify coronal jets (thus named Automated Jet Identification Algorithm - AJIA). Images in Figure~\ref{fig_examples} d) are the predictions by AJIA based on the inputs in panel a). Colors in the images denote the possibility of the corresponding pixels belonging to jets. A pixel with a value of 1 (0) means that AJIA thinks there is a 100\% (0\%) chance that this pixel belongs to a jet. In the first three rows, where true jets are present in the input images, AJIA could successfully identify almost identical jets to the ground truths. In the last two rows, where there are no jets but our traditional jet identification algorithm SAJIA \citep{Liu2023Power} wrongly detects jets, AJIA successfully avoids making the same mistakes.

\subsection{Model evaluation}

To evaluate the performance of AJIA, a threshold needs to be defined - only above which can the detected feature by AJIA be considered as a jet. For example, if we have a threshold of $T$ and a detected feature by AJIA can be considered as a jet (true/positive) only if the maximum predicted value of the feature by AJIA is no less than $T$. Otherwise, it is considered as non-jet (fake/negative). Panel a) in Figure~\ref{fig_matrix} shows how the recall, the true negative Rate (TNR), the precision, and the F-measure evolve with different thresholds $T$. Recall, TNR, precision, and F-measure are defined as follows:

\begin{equation}
\begin{split}
Recall = \frac{TP}{TP+FN}, \\
TNR = \frac{TN}{TN+FP}, \\
Precision = \frac{TP}{TP+FP}, \\
F-Measure = \frac{2 * Precision * Recall}{Precision + Recall}, 
\end{split}
\end{equation}

\noindent where $TP$ is true positive (AJIA successfully detects the jet in the label), $FN$ is false negative (AJIA misses the jet in the label), $TN$ is true negative (there is no jet in the label, and AJIA also does not detect any jet), and $FP$ is false positive (there is no jet in the label but AJIA detects a jet). One can see from the above definitions that recall represents the percentage of true jets in the labels that are successfully detected by AJIA. TNR denotes the percentage of fake jets in the labels that are also considered fake jets by AJIA. Precision is the percentage of true jets in all jets detected by AJIA. F-measure measures the combined effect of precision and recall.

We can see from Figure~\ref{fig_matrix} a) that both TNR and precision increase as the threshold grows. However, the recall decreases with the threshold. When the threshold is around 0.76, the F-measure peaks at $\sim$0.81, and all three other measurements converge at similar values. This suggests that a threshold of 0.76 would give the most ideal and balanced performance. Figure~\ref{fig_matrix} b) is a comparison between the ground truths and the predictions of AJIA. Among 1040 candidates identified by AJIA, 835 (205) are true (fake) jets, yielding a precision of $\sim$80.3\%.

\subsection{Application to higher-cadence data} \label{sec:43}

The current dataset used for building the model was generated by \cite{Liu2023Power} and \cite{Soos2024} and contains coronal jets detected with a cadence of 3 hours. This low cadence, together with the relatively low precision of SAJIA, results in a missing rate of $\sim30$\% in non-polar regions \cite[see estimations in ][]{Liu2023Power} and prevents us from further studying the temporal evolution of the detected jets. The high performance of AJIA indicated by its high recall, TNR, precision, and F-measure, as detailed in the previous subsection, provides an excellent opportunity to look into the above issue via automatically detecting jets with high accuracy at higher cadences.

To test the application of AJIA to higher-cadence data, we employed SAJIA to detect coronal jets at 1-hour intervals from 00:30 UT every day throughout January 2011. SAJIA yields 409 jet candidates, compared to 68 jet candidates given by SAJIA with a 3-hour cadence in January 2011. After laborious identification of these jet candidates by downloading and checking their temporal evolution one by one, 235 are identified as true, and the other 174 are fake. Among all fake jets, $\sim94\%$ are (part of) prominences, CMEs, or coronal rains. This gives a precision of SAJIA of $\sim51\%$, consistent with the findings in \cite{Liu2023Power} and \cite{Soos2024} (also see discussions in Sect.~\ref{sec:conc}).

\begin{figure}[tbh!]
\centering
\includegraphics[width=\hsize]{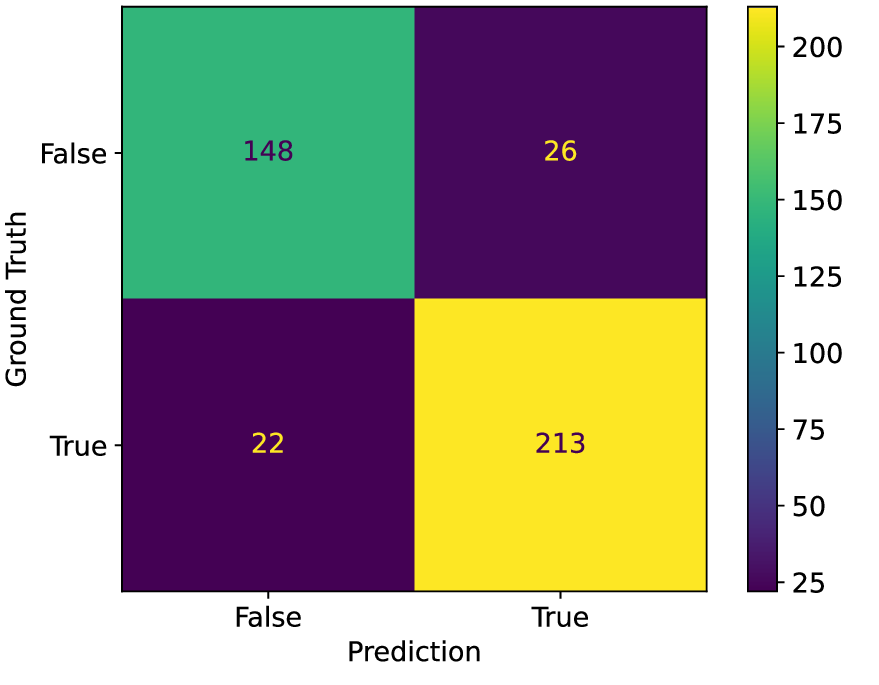}
\caption{\textbf{Confusion matrix of AJIA with 1-hour cadence data.} Similar to panel b) in Fig.~\ref{fig_matrix}, this figure shows the distribution of true and fake jets in the input labels and predictions made by AJIA from the 235 true and 174 fake jets detected in January 2011.}
\label{fig_confusion}
\end{figure}

These jets were not included in the previous dataset employed to build AJIA and could be used to test the application of AJIA to unknown events. Figure~\ref{fig_confusion} depicts the confusion matrix of AJIA's prediction on the above 235 true and 174 fake jets. TP, TN, FP, and FP are 213, 148, 26, and 22, respectively. This indicates that the precision of AJIA in detecting these unknown events is about 0.81. The recall, TNR, and F-measure are 0.81, 0.80, and 0.81, respectively. These values are consistent with what was found in the validation set as described in the previous subsections and further suggest AJIA's potential to detect off-limb coronal jets accurately.

\begin{figure}[t!]
\centering
\includegraphics[width=\hsize]{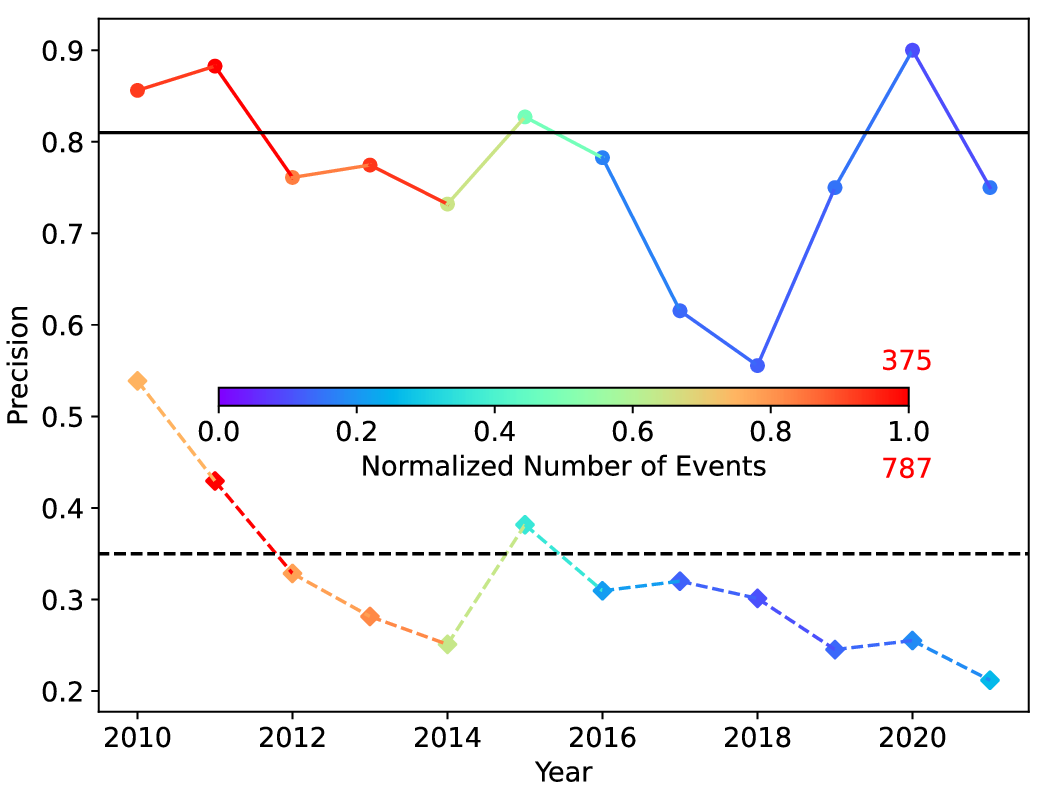}
\caption{\textbf{Precision of AJIA and SAJIA for jets in different years.} Colored dots and the solid curve are the yearly precisions of AJIA, with colors denoting the normalized (to the red number 375) number of events. Colored diamonds and the dashed curve are the yearly precisions of SAJIA, with colors denoting the normalized (to the red number 787) number of events, using data from \cite{Soos2024}.}
\label{fig_precision}
\end{figure}

\section{Conclusions and Discussions} \label{sec:conc}

In this paper, we presented the development of the Automated Jet Identification Algorithm (AJIA), which is built based on off-limb coronal jets detected by our previously developed semi-automated jet identification algorithm \citep[SAJIA,][]{Liu2023Power}. These jets were fed into a U-NET \citep{Ronneberger2015} neural network to train the final model. Evaluating AJIA on a test set containing 2120 true and fake jets yields a precision, recall, TNR, and F-measure of around 0.81, where the precision is significantly larger than that of SAJIA (0.34).

It was found in \cite{Soos2024} that the precision of SAJIA is heavily impacted by the CCD degradation of SDO/AIA. Diamonds connected by dashed lines in Figure~\ref{fig_precision} are the precisions of SAJIA measured each year \citep[inferred from Table 1 in ][]{Soos2024}, with colors denoting the normalized number of events. In general, the precision of SAJIA undergoes an overall decreasing trend, which is consistent with the overall decreasing sensitivity of the SDO/AIA detectors \citep[e.g.,][]{Santos2021}. We demonstrate that AJIA is not affected by the same effect, although it was trained using SDO/AIA 304 \AA\ images before being corrected for CCD degradation. Dots connected by solid lines in Figure~\ref{fig_precision} are the precisions of AJIA measured each year, where colors are the normalized number of events. It can be seen that the performance of AJIA does not decrease with time, and its minimum value ($>$0.55 in 2018) is above the maximum precision of SAJIA ($\sim$0.54 in 2010).

To conclude, AJIA is a step forward compared to SAJIA as it is more precise (with a precision of 0.81) and not affected by CCD degradation. AJIA is also fast, and it takes less than 6 seconds to make predictions for all 2120 images in the test set (2.6 ms per image). Another advantage of AJIA should be noted - it gives the ``possibility'' of whether a pixel in the observation belongs to a jet or not (see panel c in Fig.~\ref{fig_examples}). This enables us to generate jet heatmaps directly from SDO/AIA 304 \AA\ observations and allows real-time jet detection and visualization. These advantages of AJIA are essential in enabling many pieces of research, including but not limited to studying the collective behavior of coronal jets over the long term, their evolution over the solar activity cycles, and their relation with other solar phenomena \citep[see e.g.,][]{Liu2023Power, Soos2024}. 

Future work will also focus on improving AJIA's detection precision, which might be achieved by adding several fully connected layers after U-NET instead of giving a fixed threshold, as was done in this work.  Another future work will utilize the improved model to detect more jets at a much higher cadence (i.e., 1 hour or less) than 3 hours to explore the temporal aspects of coronal jets, especially given the capability of AJIA in accurately detecting jets from such data as described in Sect.~\ref{sec:43}. This will also enable the study of their velocities by developing a dedicated automated algorithm using, including but not limited to, the surfing transform technique \citep{Uritsky2023}, the Gaussian fitting method \citep{Chitta2023}, and the optical flow estimation \citep{Fleet2006}. Their kinetic energy would further be estimated, and the existence of a power-law distribution, which is essential in understanding the fundamental physics of the release of free magnetic energy in the solar atmosphere, would then be examined following \cite{Liu2023Power, Uritsky2023}.

This will further enable a series of statistical studies that were not done before due to the relatively small number of events detected. For example, \cite{Liu2023Power} found the ``butterfly diagram'' of coronal jets where the average latitudes of jets migrate from mid-latitudes to the equator from the beginning to the end of the solar activity cycle. It is well known that magnetic elements in high latitudes also migrate toward the polar regions throughout the solar cycle. However, this trend was not seen in \cite{Liu2023Power}, and whether its absence is caused by the limited number of events or the possibly different triggering mechanisms between jets originating from active regions and non-active regions is yet to be examined by building a larger dataset with more events.

Via having more samples of off-limb coronal jets, the distributions and differences of active-region, quiet-region, and polar jets could be further studied. Their different behaviors during solar activity could also be evaluated. The future large dataset would also enable statistical studies on how coronal jets could gain their kinetic energy \citep[e.g.,][]{Liu2014How}, how many twists they release \citep[e.g.,][]{Liu2019How}, and how the magnetic energy is distributed to different forms of energies during their eruption \citep[e.g.,][]{Liu2016b}. Moreover, preliminary evidence of the so-called solar active longitude \citep[e.g.,][]{Gyenge2017} was given in both \cite{Liu2023Power} and \cite{Soos2024}, but more evidence could be supplied by studying a significantly larger number of jets. Finding and validating active longitudes from these small-scale events will be significant for the theory and simulation of the solar dynamo.

\hspace{3cm}
\\
\\
\noindent \textbf{Acknowledgements}

\noindent {\it We acknowledge the use of the data from the Solar Dynamics Observatory (SDO). SDO is the first mission of NASA's Living With a Star (LWS) program. The SDO/AIA data are publicly available from NASA's SDO website (\url{https://sdo.gsfc.nasa.gov/data/}). This research is supported by the Strategic Priority Research Program of the Chinese Academy of Science (Grant No. XDB0560000), National Key Technologies Research, Development Program of the Ministry of
Science and Technology of China (2022YFF0711402), the Informatization Plan of the Chinese Academy of Sciences (CAS-WX2022SF-0103), and the National Natural Science Foundation (NSFC 12373056, 42188101). Yimin Wang and Ye Jiang acknowledge the support from the National Natural Science Foundation (NSFC 12303103) and the Natural Science Foundation of Shandong Province (ZR2023QF151). R. Erd{\'e}lyi is grateful to STFC (UK, grant number ST/M000826/1) and PIFI (China, grant number No. 2024PVA0043). M.B. Kors{\'o}s acknowledges support by the Leverhulme Trust Found ECF-2023-271 and UNKP-23-4-II-ELTE-107, ELTE Hungary. R. Erd{\'e}lyi and M.B. Kors{\'o}s also thank for the support received from NKFIH OTKA (Hungary, grant No. K142987). Sz.S. acknowledges the support (grant No. C1791784) provided by the Ministry of Culture and Innovation of Hungary of the National Research, Development and Innovation Fund, financed under the KDP-2021 funding scheme.}

%

\vspace{5mm}
\facilities{SDO (AIA)}

\bibliographystyle{aasjournal}

\end{document}